\documentclass[prx,reprint,showpacs,superscriptaddress,longbibliography]{revtex4-1}
\usepackage{amsmath,amssymb,bm,mathrsfs,graphicx, braket, times,amsthm,enumerate, scrextend}
\usepackage[colorlinks=true,citecolor=blue,linkcolor=blue]{hyperref}
\usepackage{longtable,float}
\usepackage{multirow}
\usepackage{wasysym}
\usepackage[all,cmtip]{xy}
\usepackage[normalem]{ulem}
\usepackage[usenames,dvipsnames]{color}

\renewcommand{\vec}[1]{\bm{#1}}

\begin{document}
\title{Fragile Topology and Wannier Obstructions}

\begin{abstract} 
Topological phases, such as Chern insulators, are defined in terms of additive indices that are stable against the addition of trivial degrees of freedom. Such topology presents an obstruction to any Wannier representation, namely, the representation of the electronic states in terms of symmetric, exponentially localized Wannier functions. Here, we address the converse question: Do obstructions to Wannier representation imply stable band topology? We answer this in the negative, pointing out that some bands can also display a distinct type of ``fragile topology." 
Bands with fragile topology do not admit any Wannier representation by themselves, but such a representation becomes possible once certain additional trivial degrees of freedom are supplied.
We construct a physical model of fragile topology on the honeycomb lattice that also helps resolve a recent puzzle in band theory. This model provides a counterexample to the assumption that splitting of an ``elementary band representation'' introduced in  [Nature 547, 298--305 (2017)]  leads to bands that are individually topological. Instead, half of the split bands of our model realizes a trivial band with exponentially localized symmetric Wannier functions, whereas the second half possess fragile topology. 
Our work highlights an important and previously overlooked connection between band structure and Wannier functions, and is expected to have far reaching consequences given the central role played by Wannier functions in the modeling of real materials.
\end{abstract}

\author{Hoi Chun Po}
\affiliation{Department of Physics, Harvard University, Cambridge MA 02138}

\author{Haruki Watanabe}
\affiliation{Department of Applied Physics, University of Tokyo, Tokyo 113-8656, Japan}

\author{Ashvin Vishwanath}
\affiliation{Department of Physics, Harvard University, Cambridge MA 02138}

\maketitle
{\it Introduction}---In recent years, there has been rapid development of our understanding of  topological phases. One important area of activity has been to classify all possible, distinct gapped phases by relating them to appropriate mathematical classification schemes.  For noninteracting fermions  in the presence of only internal symmetries (such as time reversal symmetry), a classification of states with a bulk energy gap has been obtained using K-theory \cite{Kitaev}. Conceptually, states with nontrivial topology are readily identified from their gapless edge states, while the trivial state has boundary that can be gapped without breaking symmetry \cite{Schnyder, Ryu2010, LukaszGlue, Molenkamp}. An important feature of such classification schemes is the notion of stability, that is, the requirement that a topological phase is robust to the addition of trivial, weakly coupled degrees of freedom \cite{Kitaev, Schnyder, Ryu2010, LukaszGlue, Molenkamp, Charles}.

For crystalline systems, however, new physical complications arise. First, the presence of crystalline symmetries can protect new kinds of topological phases \cite{MirrorChern,FuPRLTCI},
which are captured by extensions of the K-theory classification scheme to the case of crystalline symmetries \cite{MooreFreed, Ktheory, Combinatorics,Gomi, Ken2017}.
Unlike phases protected solely by internal symmetries, phases protected by crystalline symmetries may not possess any gapless surface states, as it may not be possible to find a surface that respects all the protecting symmetries.
Second, there are cases where, despite a clear bulk distinction between two phases, it is physically unclear which one is to be labelled trivial and which topological; rather, what is well defined is {\it relative} topology, which concerns if two systems are distinct phases separated by a bulk gap closing when symmetries are preserved throughout the deformation. 
For instance, spatial symmetries can lead to mutually distinct product states \cite{Oshikawa_SPt}. Specializing the discussion to electronic phases, product states correspond to strict atomic insulators, defined as the full filling of a set of strictly localized orbitals.
Their insulating nature can be explained even if one models electrons as classical particles trapped in periodic potential wells \cite{SciAdv}.
Such phases do not have any symmetry-protected quantum entanglement, and therefore we will label all atomic insulators as trivial \cite{Brouder, Z2Wannier,SciAdv,NC,TopoChem}.

In contrast, a weakly correlated crystal is insulating whenever the electronic energy bands, a concept defined in the momentum space, are gapped at the Fermi energy. This is more general than the real-space, atomic picture we described, and a topological band insulator arises precisely when there is an obstruction in describing the system using any atomic picture \cite{Brouder, Z2Wannier,SciAdv,NC,TopoChem}.

The contrast between the momentum- and real-space descriptions of a band insulator is closely related to the notion of Wannier functions, which is a generalization of atomic orbitals. 
Consider a group of isolated bands. Following the definitions above, we say they are {\it trivial} if and only if they can be represented in terms of exponentially localized Wannier functions that preserve all symmetries. For brevity, we will refer to this property as ``Wannier-representable,'' with the understanding that the Wannier functions involved are exponentially localized and respect all symmetries. 
In this terminology, when the set of valence bands of a system is Wannier-representable, the ground state is a trivial atomic insulators; conversely, a set of topological bands will be obstructed from any Wannier representation.

It is important to connect our discussion to the more conventional notion of (symmetry-protected) topological indices of band structures. 
Familiar topological phases characterized by, say, a nontrivial Chern number will automatically feature a Wannier obstruction \cite{Brouder}. 
More generally, such Wannier obstructions are present for any band structure with a nontrivial K-theoretic topological index \cite{Kitaev, MooreFreed, Ktheory, Combinatorics,Gomi, Ken2017}, provided that the index is {\it not} describing the mutual topology between atomic insulators \cite{Combinatorics, NC}.

However, the converse need not be true, i.e., there could be examples of bands which are {\it not} Wannier representable, even when all the K-theoretic indices are compatible with a trivial phase.
This possibility originates from a strong sense of robustness (known technically as ``stable equivalence'') demanded in the K-theory-based classification of band insulators, which is more stringent than what the physical problem calls for \cite{MooreFreed}.
We will present a concrete example in this work, where we demonstrate that a set of Wannier-obstructed bands can be explicitly trivialized simply by the presence of an additional atomic insulator. The band topology in the Wannier-obstructed bands is therefore weaker than that of familiar, stable topological phases captured within K-theory, and we will refer to it as ``fragile topology.''

We now highlight some applications of the fragile topology concept, both to the conceptual as well as to the practical aspects of band structures. 
First, we stress this concept has important implications for the modeling of  correlated materials. There, one seeks to capture the bands of interest by a tight-binding model into which one incorporates  electron-electron interactions. The tight-binding model itself is obtained from the Wannier functions, hence if the relevant bands possess fragile topology, the resulting Wannier obstruction implies that this canonical procedure will fail at the first step. It can  be remedied only by incorporating additional carefully selected orbitals. That this is not an academic issue was emphasized in recent discussions \cite{Mar2018, Jun2018, Aug2018, AhnTBG} of the band structure of twisted bilayer graphene (TBG), where  ground-breaking experiments have discovered a correlated insulator and a superconductor \cite{Kim2017,Cao2018a,Cao2018}.  It was pointed out in Ref.\ \onlinecite{Mar2018} that the relevant bands in TBG suffers from a Wannier obstruction if all symmetries are included, despite the absence of any know stable topological phase in this symmetry setting.

Second, we point out a proper understanding on fragile topology is crucial for utilizing the recent proposal in Refs.\ \onlinecite{TopoChem,Graph, DBilbao}, which asserted that an exhaustive comparison of symmetry representations in momentum space and real space will lead to an efficient diagnosis of topological materials. We clarify that, if the supposedly topological filled bands feature only fragile but not stable topology, the actual topological properties of the materials could be trivialized merely by the presence of filled bands corresponding to closed-shell electrons.
Finally, we note that although topological properties of the (non-abelian) Berry phase of band structures are commonly taken as a defining feature of topological crystalline phases \cite{Yu2011, Taherinejad2014, Alexandradinata2014, Alexandradinata2016, Holler2017}, they imply only a Wannier obstruction and not necessarily stable topology.

{\it A honeycomb lattice model}---We will begin by describing our construction of a four-band model on the honeycomb lattice, which splits into conduction and valence bands, each consisting of two bands, separated by a band gap. We will later show that the valence bands of our model are trivial, whereas the conduction bands feature fragile topology.

Consider a honeycomb lattice with the origin placed at the center of a hexagon and a $p_z$ orbital localized to each of the sites. We assume the system is symmetric under time-reversal but not spin rotations, which is a natural setting in the presence of strong spin-orbit coupling.
We will assume the spatial symmetries of the wallpaper group No.\ 17 (SG No.\ 183), which describes the symmetries of the 2D system placed on a symmetry-matched substrate \cite{Z2_Graphene,Z2_QSH,TopoChem, Graph}.
We denote this wallpaper group by $\mathcal G$, 
and let $\mathcal P $ be its point group ($6mm$). 
$\mathcal P$ is generated by a six-fold rotation about the hexagon center and a mirror along a line passing through a nearest-neighbor bond.
We will always assume periodic boundary conditions.

\begin{figure}[tbh]
\begin{center}
{\includegraphics[width= 0.48 \textwidth]{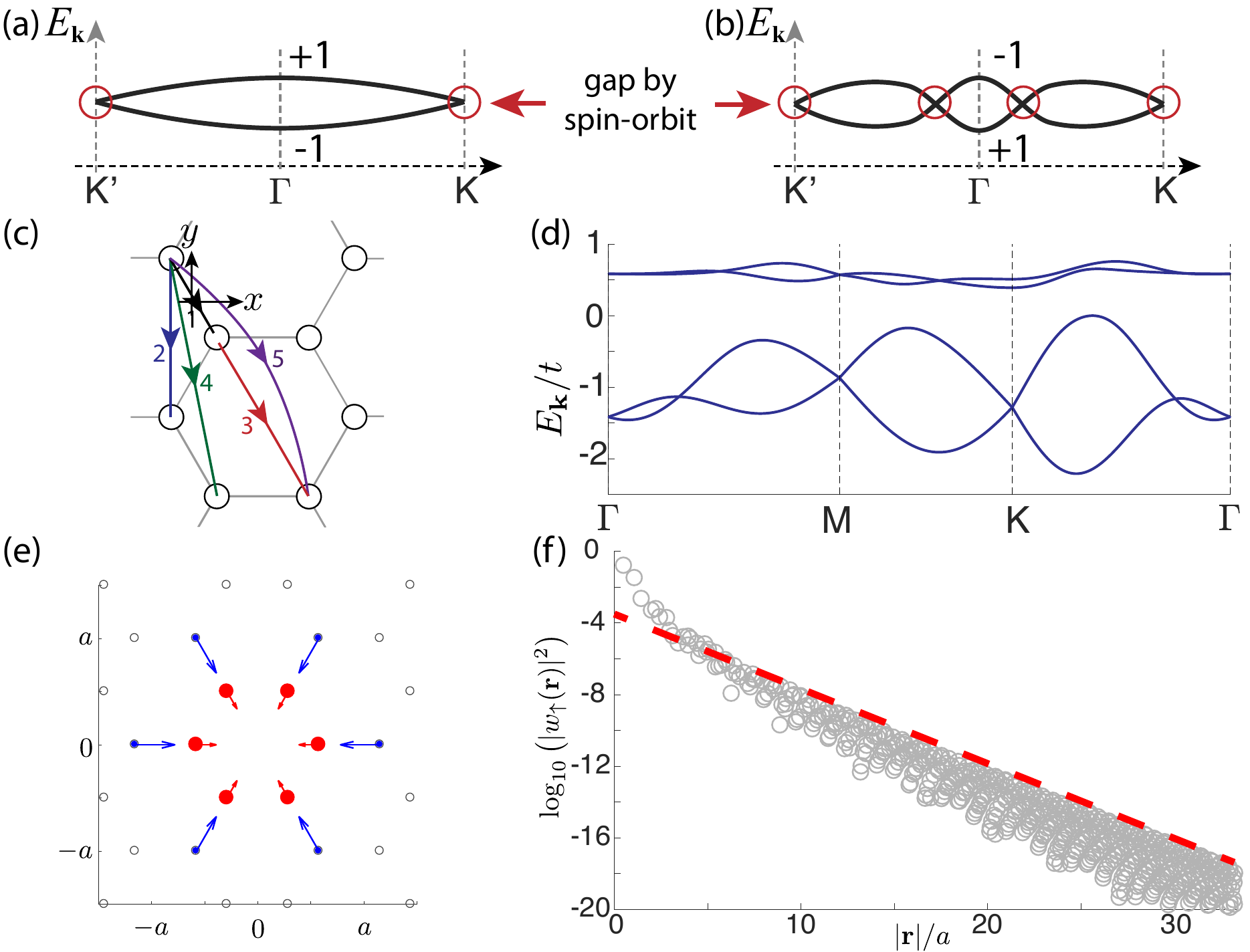}} 
\caption{
{\bf Model for fragile topology.} 
(a,b) Construction of model with fragile topology. (a) The nontrivial $\mathbb Z_2$ index of the Kane-Mele model \cite{Z2_Graphene,Z2_QSH} can be diagnosed from the inversion eigenvalues of the bands \cite{FuKane}, as we labeled for $\Gamma$. The role of spin-orbit coupling is to open a gap. (b) A further band inversion at $\Gamma$ trivializes the $\mathbb Z_2$ index. 
(c) Terms in the Hamiltonian $\hat H_0$. (d) Band structure of the honeycomb Hamiltonian, with a clear band gap at a filling of two electrons per unit cell. 
(e) A symmetric Wannier function $w_{\uparrow} (\vec r)$ centered at the origin. $a$ denotes the lattice constant.
The other Wannier function, also centered at the origin, can be obtained by applying time-reversal.
We visualize it by an arrow and a circle attached to every site at which $|w_{\uparrow} (\vec r)|^2 > 0.3 \times 10^{-2}$. Red (blue) filled circles indicates that, locally, the spin is tilting up (down), and their sizes represent the relative weights of the Wannier function. The arrow shows the in-plane component of the spin.
(f) $|w_{\uparrow} (\vec r)|^2$ decays exponentially at large distances (red dashed line).
\label{fig:WF}
 }
\end{center}
\end{figure}

Our goal is to construct a model with fragile topology. To this end, one should first rule out the presence of stable topology.
In our context, such a model can be constructed as follows: Starting with the Kane-Mele model \cite{Z2_Graphene,Z2_QSH} with inversion symmetry (Fig.\ \ref{fig:WF}a), which has a nontrivial $\mathbb Z_2$ quantum spin Hall index, we add additional terms to induce a band inversion at $\Gamma$ (Fig.\ \ref{fig:WF}b). This trivializes the $\mathbb Z_2$ quantum spin Hall index. However, as the inversion symmetry combines with a 2D $C_2$ rotation  into a mirror in the plane parallel to the system, one can only conclude that the system has an even mirror Chern number \cite{MirrorChern}. The last step, therefore, is to break inversion symmetry, and hence the mentioned mirror. This gives a model without any known topological invariant.

We now construct our model $\hat H_0$ explicitly. This is achieved by first specifying a collection of time-reversal symmetric bonds, and then symmetrizing by summing over all the $g$-related bonds for $g\in \mathcal G$. 
For the bond $i \,(= 1,2,\dots ,5) $ represented by an arrow going from site $\vec x$ to $\vec y$ (Fig.\ \ref{fig:WF}c), we define 
\begin{equation}\begin{split}\label{eq:HBond}
\hat h^{(i)} \equiv  \sum_{\alpha,\beta = \uparrow, \downarrow} \hat c_{\vec y, \alpha}^\dagger \left( \tau^{(i)} \sigma_0 + i\, \vec \lambda^{(i)} \cdot \vec \sigma \right)_{\alpha \beta} \hat c_{\vec x, \beta},
\end{split}\end{equation}
where $\hat c_{\vec x, \alpha}$ and $\hat c_{\vec y, \alpha}$ respectively denote the fermion annihilation operators for a spin-$\alpha$ electron localized at sites $\vec x$ and $\vec y$, and $\sigma_j$'s denote the standard Pauli matrices
\footnote{
Here, we take spin-up to be aligned with the crystallographic $z$ axis, which is normal to the 2D system.
}
. Here, $(\tau^{(i)}, \lambda_{1}^{(i)}, \lambda_{2}^{(i)}, \lambda_{3}^{(i)} )$ are four dimensionless real parameters defining the electron hopping along bond $i$. Their chosen numerical values are tabulated in Appendix \ref{app:details}.
Note that we have optimized our model for achieving a more sizable band gap, which leads to longer range hopping (up to fifth-nearest neighbors). Nonetheless, we stress that only finite-range hoppings are considered, and so long as both the  band gap and symmetries are maintained the general features we describe below will persist against perturbations.

The honeycomb model is then defined by 
\begin{equation}\begin{split}\label{eq:H0}
\hat H_0 = \frac{t}{|\mathcal P| } \sum_{i=1}^5 \sum_{g \in \mathcal G } \hat g \, \hat h^{(i)} \hat g^{-1}  + {\rm h.c.},
\end{split}\end{equation}
where $|\mathcal P| = 12$, and $t>0$ is an overall energy scale.
As shown in Fig.\ \ref{fig:WF}d, a gap at half filling is found at all high-symmetry momenta. Considering also the interior of the Brillouin zone (BZ), one finds a band gap \footnote{
Defined in this work as the difference between the bottom of the conduction band and the top of the valence bands, which lower bounds the ``direct band gap'' definition common in the study of topological band insulators.
}
of $0.39 t$. For comparison, the valence bandwidth is $\sim 2 t$.

Our next goal is to analyze the band topology  present in $\hat H_0$. 
In particular, we will first establish that the valence bands are trivial.

{\it Trivial valence bands}---As a first check, we construct symmetric Wannier functions of the valence bands \cite{VanderbiltRMP, Sakuma}. No obstruction was encountered in this process Appendix \ref{app:WF}, and the results are visualized in Fig.\ \ref{fig:WF}e. The weight of a Wannier function as a function of $r$, the distance away from its charge center (which sits at the center of a hexagon), is shown in Fig.\ \ref{fig:WF}f.
The weight decays exponentially as $r \rightarrow \infty$, decaying by 4 orders of magnitude in $10$ lattice constants. This implies the valence bands of $\hat H_0$ admit symmetric, exponentially localized Wannier functions, and therefore the corresponding band insulator is trivial.
In Appendix \ref{app:Deform}, we also provide an alternative proof of its triviality through an adiabatic, symmetric deformation to an explicit atomic limit.

{\it Nontrivial conduction bands}---Since the conduction bands of $\hat H_0$ combine with the valence bands to form an atomic insulator (namely, the $p_z$ orbitals localized to the honeycomb sites), the triviality of the valence bands rules out any stable topology in the conduction bands. 
However, based solely on the symmetry representations of the conduction bands (available, e.g., on the Bilbao crystallographic server \cite{Graph}), one can see that the conduction bands cannot be Wannier representable. 
This is because any atomic insulator at a filling of two electrons per unit cell must have the electrons localized to the triangular lattice sites, and one can check that no such atomic insulator possesses the same set of symmetry representations as the conduction bands at all momenta \cite{TopoChem,Graph,DBilbao}. 
As symmetry representations are bulk quantities which cannot be modified without a band-gap closing, they serve as a nontrivial topological invariant for the conduction bands.
Note that a representation-based invariant is as robust as that arising from more involved objects, say from the Wilson loops \cite{Yu2011, Taherinejad2014, Alexandradinata2014, Alexandradinata2016, Holler2017}.

It is conceptually revealing to connect our present observations to the discussions in Ref.\ \onlinecite{NC}, which discussed how symmetry eigenvalues can inform band topology in two different ways. The first is embodied in the notion of ``symmetry-based indicators of band topology,'' which is concerned with stable topology.
The second diagnosis is more subtle, and is tied to the distinction between the physical stacking and the formal addition of atomic insulators, where a (formal) subtraction between atomic insulators is allowed only in the latter but not the former. 
By formal subtraction, we mean the following: let $\mathcal A$, $\mathcal V$ and $\mathcal C$ be gapped bands such that $\mathcal A$ can be viewed as arising from stacking the bands $\mathcal V$ and $\mathcal C$, i.e., $\mathcal A = \mathcal V \oplus \mathcal C$.
Then we can formally identify $\mathcal C$ as the difference between $\mathcal A$ and $V$.
Crucially, even when both $\mathcal A$ and $\mathcal V$ above are atomic insulators, $\mathcal C$ may not admit a Wannier representation.
Such is the case for the conduction bands ($\mathcal C$) of $\hat H_0$, which can be viewed as subtracting the valence bands ($\mathcal V$), an atomic insulator with electrons localized to the triangular sites, from the atomic insulator formed by full filling of the honeycomb sites ($\mathcal A$).
In fact, such systems are prime candidates for fragile topology, since by definition the symmetry representations do not indicate any necessary stable topology. 
We also note that some ``filling-enforced quantum band insulators'' discussed in Ref.\ \onlinecite{SciAdv} also sit in this class \cite{NC}, and therefore they might be early examples of fragile topology.

{\it Fragile topology and band representations}---Having established the existence of fragile topology in the conduction bands of the concrete model $\hat H_0$, we now discuss its general implications. 
As we have alluded to, the key difference between stable and fragile topology descends from the notion of ``stable equivalence'' in K-theory (Fig.\ \ref{fig:LEGOs}a; also see Appendix \ref{app:K}). It is instructive to provide a more precise definition.
Consider an isolated set of bands, and we ask if it can be represented in terms of exponentially localized Wannier functions that respect all symmetries. If this is possible, the set of bands is trivial. If this fails, we can further ask: can we add to this set another set of trivial bands, derived from an atomic insulator, and then obtain localized Wannier functions? If yes, our original set only possesses fragile topology (Fig.\ \ref{fig:LEGOs}b,c). By this definition, the valence bands of our model are trivial, and the conduction bands possess fragile topology. We also provide a more physical perspective on the preceding discussion in Appendix \ref{app:PhysView}.

\begin{figure}[h]
\begin{center}
{\includegraphics[width=0.48 \textwidth]{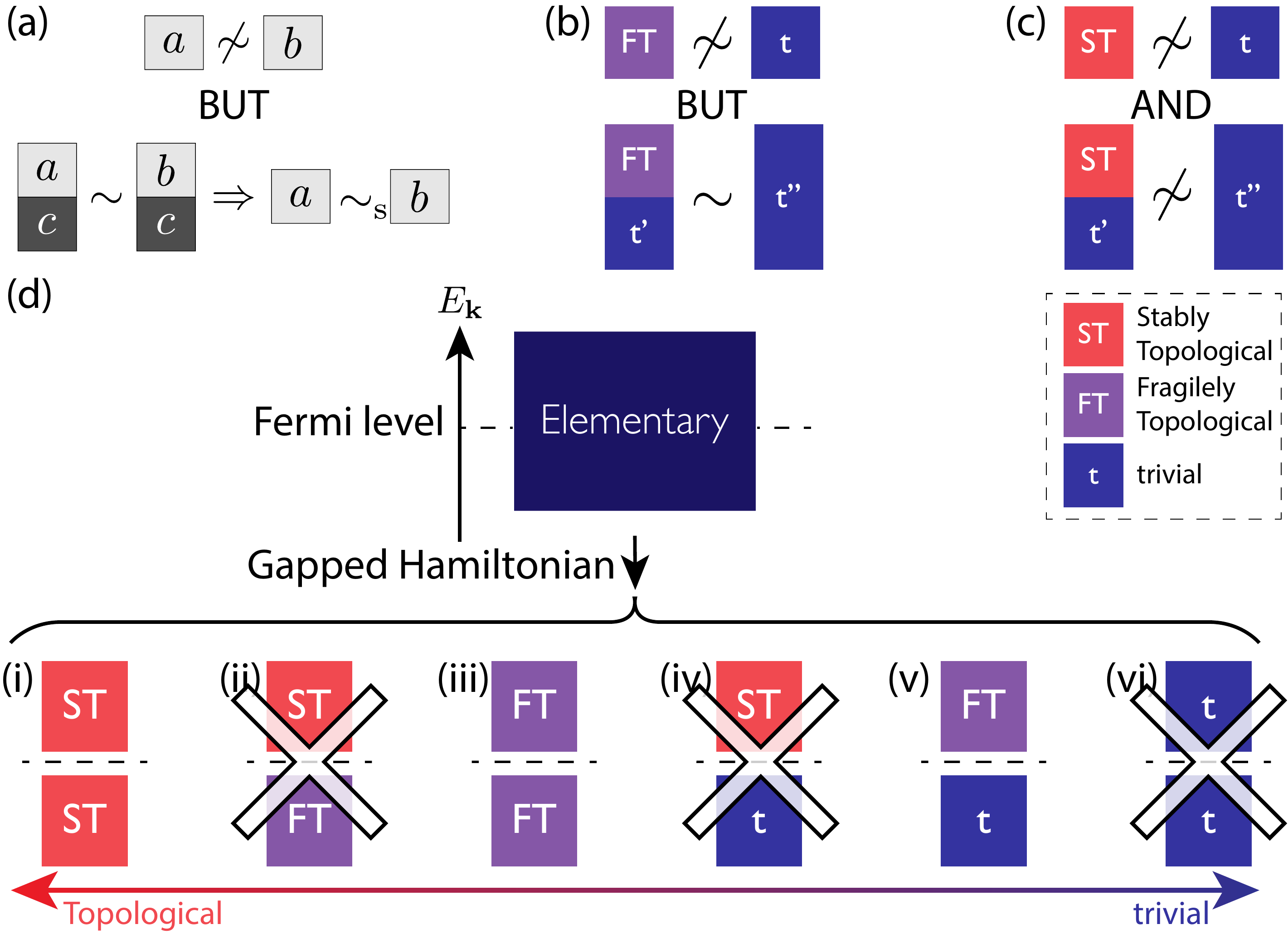}} 
\caption{{\bf Fragile topology and decomposable elementary band representations.} (a) In a K-theory-based classification\cite{Kitaev, MooreFreed, Ktheory, Combinatorics,Gomi, Ken2017}, it could be that two band insulators $a$ and $b$ are not smoothly connected, but the obstruction is resolved once we append to both sides an additional set of bands $c$. We say $a$ and $b$ are stably equivalent, denoted by $a \sim_{\rm s} b$. 
(b) Using similar ideas, we introduce the notion of ``fragile topology.'' We say a set of bands possesses fragile topology if there is a topological obstruction in deforming it to any trivial (atomic) limit, but the obstruction is resolved once we allow for the introduction of additional trivial degrees of freedom.
(c) In contrast, bands with stable topology remain nontrivial upon the stacking of any trivial bands.
(d) Up to an overall sign change of the tight-binding Hamiltonian, there are generally six distinct splitting patterns for a decomposable elementary band representation. Three cases are ruled out by definitions, indicated by crosses. Our model $\hat H_0$ shows that case (v), the only splitting pattern with trivial valence bands, is  possible.
\label{fig:LEGOs}
 }
\end{center}
\end{figure}

Next, we connect the phenomenology of fragile topology to a recent proposal \cite{TopoChem,Graph, DBilbao} that the theory of band representations, developed by Bacry, Michel, and Zak \cite{Zak1982, Bacry1988, Zak2000, Zak2001}, can lead to a general classification of topological band insulators. (Please see Appendix \ref{app:BR}  for a brief introduction to the notion of ``band representations.'')
A key idea in this proposal is that of an ``elementary band representation'' (EBR) \cite{Zak1982, Bacry1988, Zak2000, Zak2001}, which has the defining feature that, if it splits in momentum space into disconnected (i.e., separated by a band gap) conduction and valence bands, then the two sets of bands cannot be simultaneously trivial \cite{TopoChem,Graph, DBilbao}.

In fact, as is discussed in depth in Refs.\ \onlinecite{TopoChem,Graph}, the tight-binding degrees of freedom specified in $\hat H_0$ correspond to an EBR, and therefore $\hat H_0$ serves as a concrete model realization of the splitting pattern ``EBR = trivial $\oplus$ fragilely topological.''
Given the ground state of $\hat H_0$ is an atomic insulator, we conclude that the splitting of an EBR does not generally imply nontrivial band topology in an insulator.

{\it Discussion}---We emphasize that while  fragile topology may be reminiscent of the ``Hopf insulator'' \cite{Hopf, HopfSurface}, they are sharply distinct concepts. Without adding additional symmetries, the Hopf insulator is only topological  in a half-filled two-band model \cite{HopfNew}, and is unstable even against the introduction of high-energy, unoccupied degrees of freedom. Therefore, topology of the Hopf insulator variety is not expected to be relevant to electronic band structures in materials, where such high energy bands are inevitably present. In contrast, fragile topology remains well-defined in this setting, 
and are only trivialized by adding fully filled atomic bands {\it below} the Fermi level. The distinction between adding orbitals above and below the Fermi energy is important because only the latter necessitates a modification of the many-body wave function. In addition, unlike the Hopf insulator, the notion of fragile topology does not rely on the nontriviality of any special map between spaces, and is ultimately defined via the symmetry protection of certain patterns of quantum entanglement in the ground state wave function. It is expected to arise whenever the spatial symmetries are rich enough, for systems with or without spin-orbit coupling and/ or time-reversal invariance, in both two and three dimensions.

Our model also proves that the splitting of an EBR, discussed in Refs.\ \onlinecite{TopoChem,Graph, DBilbao}, does not necessarily imply a topological band insulator, and for a reliable diagnosis one must further corroborate analysis using symmetry eigenvalues \cite{Ari10,  Ari, Bernevig, Chern_Rotation,NC,TopoChem, Graph, DBilbao, MSG} or, more generally, wave-function-based topological invariants. 
We note that the range of physical signatures that a fragile topological phase can exhibit is expected to be restricted by the fact that it can be trivialized by stacking with an atomic insulator.
For instance, fragile topological phases are not expected to host protected surface states, for their bulk topology can be trivialized simply by the addition of atomic insulators without any surface signatures. 
This also suggests fragile topology is not expected when only internal symmetries are present, since one can always find a surface which preserves all internal symmetries.
We leave the analysis of their physical signatures and general diagnosis to future works.

In closing, we emphasize that the notion of fragile topology will be important in the modeling of electronic systems, given the central role played by Wannier functions in the well-established methods.  Moreover, fragile topology is not a mere mathematical possibility, but arises in realistic models and potentially even in real materials, like small-angle twisted bilayer graphene \cite{Mar2018, Jun2018, Aug2018, AhnTBG}.
\\

\begin{acknowledgments}
We acknowledge A.\ Akhmerov, B.\ A.\ Bernevig, B.\ Bradlyn, J.\ Kruthoff, R.-J.\ Slager, D.\ Vanderbilt, D.\ Varjas,  J.\ van Wezel, and M.\ Zaletel for discussions. 
AV and HCP were supported by NSF DMR-1411343. HW acknowledges support from JSPS KAKENHI Grant Number JP17K17678.
\end{acknowledgments}

\bibliography{references}

\clearpage

\appendix
\section{Further details of the model
\label{app:details}}

\begin{table}[H]
\begin{center}
\caption{{\bf Numerical parameters in the model.} 
Parameterized according to Eq.\ (1) in the main text.
Zeroes are omitted from the table for clarity.
\label{tab:HPara}}
\begin{tabular}{c|cccc}
Bond $(i)$& $\tau^{(i)}$ & $\lambda_1^{(i)}$ & $\lambda_2^{(i)}$ & $\lambda_3^{(i)}$\\
\hline
$1$ & $-0.7$ & $-0.4$ & $-0.2$ & ~\\
$2$ & ~ & $-0.6$ & ~ & $-1.0$\\
$3$ & $-0.3$ & $-0.7$ & $-0.4$ & ~\\
$4$ & ~ & $0.9$ & $0.3$ & ~\\
$5$ & $-0.2$ & $0.4$ & $0.3$ & ~\\
\hline \hline
$6$ & $-3.0$ & $-2.1$ & $1.2$ & ~\\
$7$ & $0.5$ & $1.2$ & $-0.7$ & ~
\end{tabular}
\end{center}
\end{table}

While Eqs.\ (1) and (2) in the main text, together with the numerical parameters in Table \ref{tab:HPara}, fully determine our model, for readers' convenience we present below an explicit matrix form of $\hat H_0$. To achieve that, we first specify our coordinate system. Let $a$ be the lattice constant, and let $\{ \vec a_i \}$ and $\{ \vec b_i \}$ respectively be the primitive lattice vectors and the corresponding reciprocal lattice vectors. We pick
\begin{equation}\begin{split}\label{eq:}
\vec a_1 =& a \left( \frac{\sqrt{3}}{2} \hat x - \frac{1}{2} \hat y\right); ~~~ \vec a_2 = a\,  \hat y \, ;\\
\vec b_1 =& \frac{4\pi}{ \sqrt{3} a}  \hat x; ~~~ \vec b_2 =  \frac{4\pi}{ \sqrt{3} a}  \left( \frac{1}{2} \hat x + \frac{\sqrt{3}}{2} \hat y\right).
\end{split}\end{equation}
The honeycomb sites are at $\vec r_1 = \vec a_1/3 + 2 \vec a_2/3$ and  $\vec r_2 = 2 \vec a_1/3 + \vec a_2/3$ (matrix indexing of the sites proceeds in that order). We parameterize the BZ by $\vec k = (g_1 \vec b_1 + g_2 \vec b_2)/2\pi $, where $g_i \in (-\pi, \pi]$.

We expand the Bloch Hamiltonian for $\hat H_0$ as 
\begin{equation}\begin{split}\label{eq:}
H_0 (g_1,g_2) = \frac{t}{12} \sum_{i,j=0}^3 h_{ij}(g_1,g_2)\,  \sigma_i \otimes \sigma_j,
\end{split}\end{equation}
where the first set of Pauli matrices corresponds to the site degrees of freedom, the second set is for the spin, and
\onecolumngrid
\begin{equation*}\begin{split}\label{eq:}
h_{00} = & -\frac{4}{5} (\cos (g_1 -g_2 )+\cos (2 g_1 +g_2 )+\cos (g_1 +2 g_2 )); \\
h_{01} = & \frac{3}{5} \left(2 ( \sin (g_1 +g_2 )-\sin g_1)-\left(4+\sqrt{3}\right) \sin \left(\frac{3 g_2 }{2}\right) \cos \left(g_1 +\frac{g_2 }{2}\right)+4 \sin g_2 \right); \\
h_{02} = & \frac{1}{5} \sin \left(g_1 +\frac{g_2 }{2}\right) \left(\left(6+8 \sqrt{3}\right) \left(\cos \left(g_1 +\frac{g_2 }{2}\right)+\cos \left(\frac{g_2 }{2}\right) \cos g_2 \right)-\left(3+16 \sqrt{3}\right) \cos \left(\frac{g_2 }{2}\right)\right); \\
h_{10} = & \frac{2}{5} (3 \sin g_1  \sin g_2 -9 \cos g_1  \cos g_2 -7 (\cos g_1 +\cos g_2 +1)); \\
h_{11} = & 
\frac{  3-\sqrt{3}  }{10}(3 \sin (2 g_1 )-2 \sin (g_2 )-3 \sin (2 g_1 +g_2 ))
 +\frac{21+4 \sqrt{3}}{5} \cos (g_1 ) \sin (g_2 )
 -\frac{3(3+\sqrt{3})}{5}  \sin (g_2 ) \cos (g_1 +g_2 )\\
&~~~ -\frac{9}{5} \sin (2 g_2 ); \\
h_{12} = & 
\frac{3 \left(1+3 \sqrt{3}\right)}{10} \left( \sin (2 g_1 +g_2 )+ \sin (2 g_1 ) \right)
+\frac{3 \left(3 \sqrt{3}-1\right)}{10} \sin (g_1 +2 g_2 )
- \frac{ 4+7 \sqrt{3}  }{5} (2 \sin g_1  \cos g_2 +\cos g_1  \sin g_2 )\\
&~~~+ \frac{\sqrt{3}-7 }{10} \sin g_1 
- \frac{2 \left(2+\sqrt{3}\right)}{5} \sin g_2 
+ \frac{3}{5} \sin (2 g_2 );  \\
h_{20} = & \frac{2}{5} (7 (\sin g_2 -\sin g_1 )-3 \sin (g_1 -g_2 )); \\
h_{21} = & \frac{1}{10} \sin g_2  \left(6 \left(3+\sqrt{3}\right) \sin (g_1 +g_2 )+\left(42+8 \sqrt{3}\right) \sin g_1 +3 \left(\sqrt{3}-3\right) \sin (2 g_1 )\right)\\
&~~~+\frac{1}{5} \sin ^2\left(\frac{g_2 }{2}\right) \left(\left(\sqrt{3}-3\right) (3 \cos (2 g_1 )-2)+36 \cos g_2 \right); \\
h_{22} = & -\frac{1}{5} \cos ^2\left(\frac{g_2 }{2}\right) \left(\left(3+9 \sqrt{3}\right) \cos (2 g_1 )+4 \left(\sqrt{3}+2 -3 \cos g_2 \right)\right)+\frac{1}{5} \sin g_1  \sin g_2  \left(\left(3-9 \sqrt{3}\right) \cos g_2 +7 \sqrt{3}+4\right)\\
&~~~+\frac{1}{10} \cos g_1  \left(6 \left(1+3 \sqrt{3}\right) \sin g_1  \sin g_2 +\left(9 \sqrt{3}-3\right) \cos (2 g_2 )+17 \sqrt{3}+1\right); \\
h_{33} = & 4 (\sin g_1 +\sin g_2 -\sin (g_1 +g_2 )).
\end{split}\end{equation*}
Note that the terms $h_{03}, h_{13}, h_{23}, h_{30}, h_{31}, h_{32}$ vanish, and are therefore omitted from the above.

We also write out the coupling Hamiltonian $\hat H_{\rm c}$ in an explicit single-particle matrix form:
\begin{equation}\begin{split}\label{eq:}
H_{\rm c}(g_1, g_2) = \frac{c}{12}
\left(
\begin{array}{ccc}
0 & 0 & C_{13} \\
0 & 0 & C_{23} \\
C_{13}^\dagger   & C_{23}^\dagger   & 0
\end{array}
\right).
\end{split}\end{equation}
Note that $H_c$ is written in a block matrix form, with the index ordering corresponding to the following basis vectors in the unit cell: $\vec r_1$, $\vec r_2$ and $\vec r_3 \equiv \vec 0$.
The $2\times 2$ sub-matrices $C_{13}$ and $C_{23}$ are functions of $(g_1,g_2)$, given by
\begin{equation*}\begin{split}\label{eq:}
C_{13}=&
\left [ 
2 \cos (g_1 )-6 \left(e^{i (g_1 +g_2 )}+e^{i g_2 }+1\right)+e^{i (g_1 +2 g_2 )} 
 \right] \sigma_0 \\
&~~~
+ \frac{1}{20}  \left[  
3 \left(21+4 \sqrt{3}\right) (\sin (g_2 )-i \cos (g_2 )+i)-2 \left(36+7 \sqrt{3}\right) e^{i (g_1 +g_2 )} \sin (g_2 ) 
 \right] \sigma_1\\
&~~~
-\frac{1}{20} i e^{-i g_1 } \left[ 
\left(7+12 \sqrt{3}\right) \left(e^{2 i (g_1 +g_2 )}+e^{2 i g_1 }-2\right)
+3 \left(4+7 \sqrt{3}\right) e^{i g_1 } \left(1+\left(1-2 e^{i g_1 }\right) e^{i g_2 }\right) 
\right] \sigma_2;\\
C_{23}=&
\left[2 \cos (g_2 ) -6 \left(e^{i (g_1 +g_2 )}+e^{i g_1 }+1\right)+e^{2 i g_1 +i g_2 } \right] \sigma_0\\
&+ \frac{1}{20} i e^{-i g_2 } \left[ \left(36+7 \sqrt{3}\right) \left(e^{2 i g_2 } -1 \right)
+3 \left(21+4 \sqrt{3}\right) \left(1-e^{i g_2 }\right) e^{i (g_1 +g_2 )}  \right]\sigma_1\\
& + \frac{1}{20} i e^{-i g_2 }  \left[  3 \left(4+7 \sqrt{3}\right) e^{i g_2 } \left(e^{i (g_1 +g_2 )}+e^{i g_1 }-2\right)
+\left(7+12 \sqrt{3}\right) \left(1+\left(1 - 2 e^{2 i g_1 }\right) e^{2 i g_2 }\right)  \right]\sigma_2 .
\end{split}\end{equation*}
One can check that the given Hamiltonians are indeed symmetric under the stated symmetries following the procedure delineated in, e.g., Ref.\ \onlinecite{NC}.\\

\twocolumngrid

\section{Constructing Wannier functions
\label{app:WF}}
We will use the ``projection method'' to construct symmetric, well-localized Wannier functions \cite{VanderbiltRMP, Sakuma}. The method proceeds by first specifying a collection of well-localized, symmetric wave functions  in real space, which serves as a reasonable guess for the actual Wannier functions. These trial wave functions are then projected into the valence-band subspace to determine a collection of unitaries which correspond to a ``smooth gauge'' for forming well-localized Wannier functions. In this process, at every $\vec k$ we have a Hermitian matrix $S_{\vec k}$ quantifying the overlap between our initial seeds and the actual valence bands. The projection procedure is well-defined and gives well-localized Wannier functions when $\det S_{\vec k} >0~\forall \vec k$. Failure of which for all choices of symmetric, well-localized trial wave functions is generally believed to be a sign of a topological band structure \cite{Brouder, Z2Wannier}. In practice, one keeps track of the stability of the procedure by monitoring the size of $\det S_{\vec k} $ across the BZ, which could be achieved by ensuring the ratio $\max_{\vec k}\{ \det S_k \} / \min_{\vec k}\{ \det S_k \}$ does not diverge \cite{VanderbiltRMP}. In addition, using the ideas in Ref.\ \onlinecite{Sakuma}, one can verify that as long as $\det S_k>0$ for all $\vec k$, the resulting Wannier functions will inherit the symmetry properties of the trial ones.

For our problem, we will construct the symmetric Wannier functions of $\hat H_0$ using the smooth deformation $\{ \hat H (\mu) \, : \, \mu \in [0,1]\}$ we discussed. Of course, the existence of symmetric, exponentially localized Wannier function is independent of the method one uses to find them, so if one prefers, one can also construct the desired Wannier functions for $\hat H_0$ without ever introducing the additional triangular sites. 
Our construction proceeds as follows: 
We start with the Wannier functions for $\hat H(1)$, which are simply the $p_z$ orbitals localized to the triangular sites. We then use these as the trial wave functions to find the Wannier functions of $\hat H(\mu')$ for some $\mu' <1$. These new Wannier functions are then used as the trials for some $\mu'' < \mu'$. We do this iteratively until we arrive at $\mu=0$, resulting in the Wannier functions for the pure honeycomb model $\hat H_0$. 
Numerically, we simply perform this procedure for the values of $\mu$ indicated by red squares in Fig.\ \ref{fig:Deform}b. 
The Wannier functions are computed using a $200\times 200$ regular momentum mesh for the BZ. On top of that, we evaluate $S_{\vec k}$ for an additional $3 \times 10^4$ randomly sampled momenta.
As $\mu$ parameterizes a symmetric smooth deformation of the ground-state wave function, we expect the projection to proceed without obstruction at every step in the construction. 
This can be verified from the behavior of $\det S_{\vec k}$ along the projection, as tabulated in Table \ref{tab:Sk}.

Note that no obstruction is encountered as we successively project the initial, tightly localized trial wave functions (using a finite number of steps) to obtain the symmetric Wannier functions of the honeycomb model $\hat H_0$. This is strongly indicative that the our Wannier functions are exponentially localized. We will not attempt to prove this in an analytic manner here; rather, we simply point to the numerical evidence that the found Wannier functions decay with an envelope which is clearly exponentially decaying.

Finally, we comment that spin-orbit coupling plays a crucial role in the existence of the symmetric, exponentially localized Wannier functions. This is reflected in a nontrivial phase winding of the wave function, locked to the in-plane spin component, as one circles the charge center. One can readily derive the required locking patterns by, e.g., writing down wave functions which are compactly supported on the six vertices of a hexagon and transforming in the stipulated way. Of course, such a wave function does not automatically satisfy the orthogonality condition with its translation copies, and therefore cannot be immediately interpreted as the Wannier functions of some parent Hamiltonian. Nonetheless, it captures the essence of the symmetry properties required, and, if preferred, one can as well use it as a trial wave function for finding the Wannier functions \cite{VanderbiltRMP, Sakuma} of the valence bands of $\hat H_0$.

\begin{table}[H]
\begin{center}
\caption{{\bf 
Data on the construction of Wannier functions through successive projections along the smooth deformation parameterized by $\mu$.} We follow the notation in Ref.\ \onlinecite{VanderbiltRMP} for the overlap $S_{\vec k}$, the gauge-invariant spread functional $\Omega_{\rm I}$, and the gauge-dependent one $\tilde \Omega$. $a$ denotes the lattice constant.
\label{tab:Sk}}
\begin{tabular}{c|cccc}
$\mu$ & $0.75$ & $0.5$ & $0.25$ & $0.00$\\
\hline
$ \min_{\vec k} \{ \det S_{\vec k} \} $ & 0.27 & 0.85& 0.83& 0.74  \\
$ \max_{\vec k} \{ \det S_{\vec k} \}/ \min_{\vec k} \{ \det S_{\vec k} \} $ & 1.53 & 1.09& 1.08 & 1.05 \\
$\Omega_{\rm I}/a^2  $& 0.38 & 0.68& 1.01& 1.20  \\
$\tilde \Omega/a^2   $ & 0.00 & 0.01& 0.02& 0.03 
\end{tabular}
\end{center}
\end{table}

\section{Deformation to an explicit atomic limit
\label{app:Deform}}
\begin{figure}[tbh]
\begin{center}
{\includegraphics[width= 0.45 \textwidth]{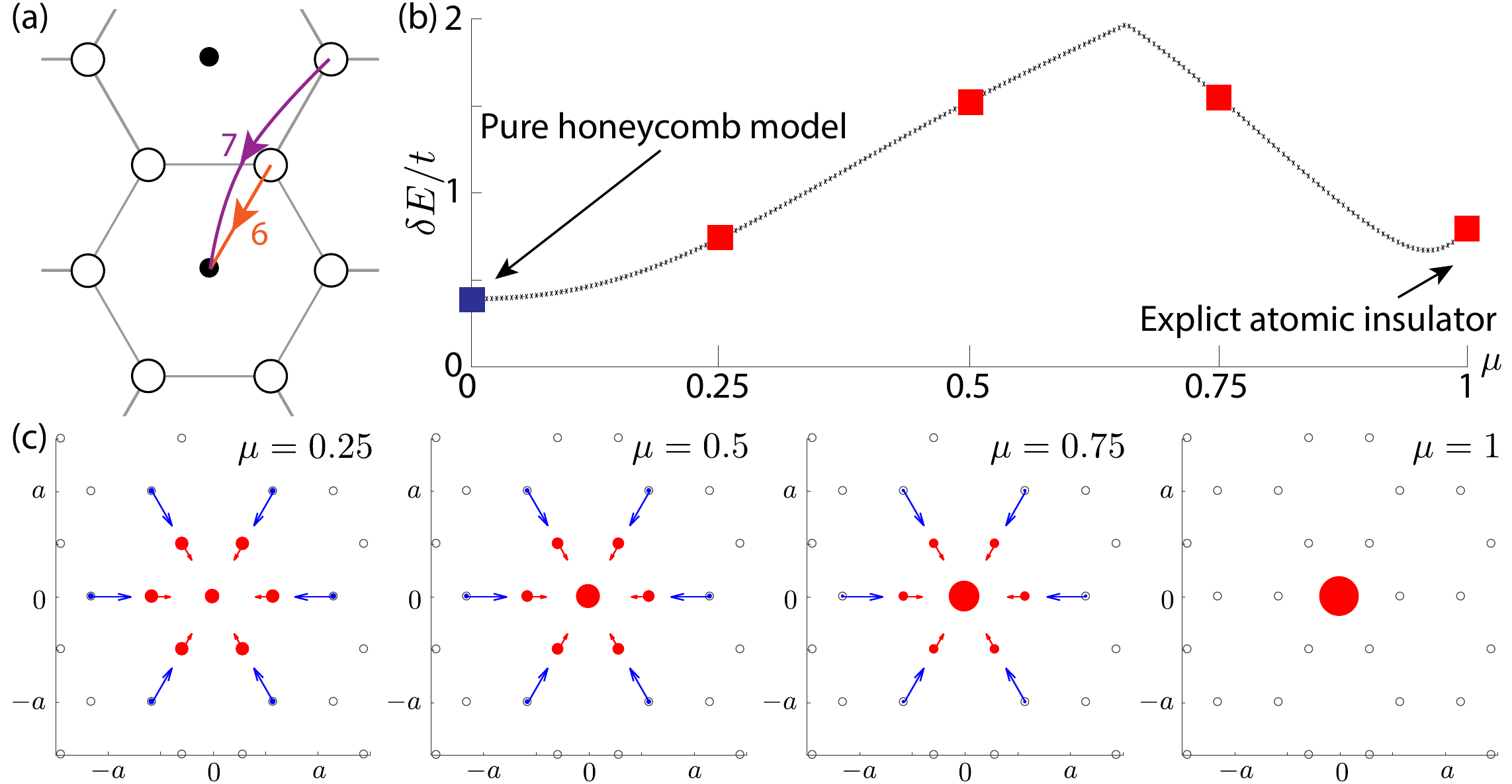}} 
\caption{
{\bf Adiabatic deformation to an atomic limit.} (a) Coupling between the honeycomb sites and additional sites at the centers of the hexagons. 
(b) Symmetric smooth deformation of the pure honeycomb ground state ($\mu = 0$) to an explicit atomic insulator with electrons localized to the centers of the hexagons ($\mu =1$). $\delta E$ denotes the band gap at filling $2$. A cross indicates a data point computed using $\sim 1350$ points along the path in Fig.\ 1b in the main text together with an additional $10^5$ random points in the Brillouin zone.
(c) Smooth evolution of the Wannier functions for $\mu$ indicated by the four red squares in (b). See Fig.\ 1c in the main text for the Wannier function in the pure honeycomb limit (blue square), and a description on how it is visualized.
\label{fig:Deform}
 }
\end{center}
\end{figure}

While the valence bands of $\hat H_0$ possess a full set of symmetric, exponentially localized Wannier functions---a property of the honeycomb model without the need for any additional degrees of freedom---a further check on their triviality concerns its deformability to an explicit atomic limit. To this end, we augment the electronic degrees of freedom in the model by an additional $p_z$ orbital localized to each of the centers of the hexagons. These additional sites form a triangular lattice by themselves. 
As the problem is set up, initially, the wave functions of the valence bands of $\hat H_0$ have zero amplitude on these new sites.  We then consider a Hamiltonian $\hat H_c$ which couples the two sets of lattice sites (Fig.\ \ref{fig:Deform}a). Combining $\hat H_c$ and $\hat H_0$ with a tunable relative strength between the two, we consider a continuous family of Hamiltonians $\{ \hat H (\mu) \, : \, \mu \in [0,1] \}$ with the following properties: 
(i) All the stated symmetries are maintained for all $\mu$;
(ii) The honeycomb and triangular lattices are decoupled for $\mu=0$ or $1$;
(iii) the two lowest bands of $\mu= 0$ are identical to those of $\hat H_0$; and 
(iv) the two lowest bands of $\mu = 1$ arise solely from the triangular lattice sites. 
$\hat H (\mu)$ therefore interpolates between $\hat H_0$ and an explicit, strongly localized atomic limit in a symmetric manner.

Next, we construct $\hat H_c$ and $\hat H (\mu) $ explicitly. The coupling Hamiltonian $\hat H_c$  is defined in the same way as $\hat H_0$ in Eqs.\ (1) and (2) in the main text, except that we sum over bonds $i = 6,7$, and replace the energy scale $t$ by a $\mu$-dependent coupling constant $c(\mu)  = t\, \cos \left[ ( 1-2\mu) \pi /2 \right]$. 
Note that $c(0) = c(1) = 0$.
In addition, we add an additional chemical potential offset (measured with respect to the Fermi energy of $\hat H_0$) between the honeycomb and triangular lattices, captured by $\hat H_{\triangle}(\mu) = 3 (1-2\mu) t \hat N_{\triangle} $,
where $\hat N_{\triangle} $ is the total electron number operator for all sites in the triangular lattice. 
$\hat H_{\triangle}(\mu)$ is designed such that the lowest two bands will be of the honeycomb (triangular) character when $\mu=0$ ($\mu=1$).
The interpolating Hamiltonian is then defined as $\hat H(\mu) \equiv \hat H_0 + \hat H_{c(\mu)} + \hat H_{\triangle} (\mu)$.

The band gap of $\hat H(\mu)$ at filling $2$ is shown in Fig.\ \ref{fig:Deform}b. The  gap never vanishes, and attains a minimum in the purely honeycomb limit ($\mu =0$). 
This corresponds to a symmetric, adiabatic deformation between the ground state of $\hat H_0$ and an explicit atomic insulator. We further visualize the Wannier functions as a function of $\mu$ in  Fig.\ \ref{fig:Deform}c, showing their expected smooth evolution.

\section{K-theory and fragile topology
\label{app:K}}
Here, we provide a slightly more formal discussion on how the notion of fragile topology is inspired by the notion of stable equivalence in K-theory. 
To this end, we first introduce some ideas from the K-theory-based classifications of band insulators \cite{Kitaev, MooreFreed, Ktheory, Combinatorics,Gomi, Ken2017}. 
Note that the following serves only as a brief introduction, and we refer the readers to the references above for a mathematically precise discussion.

Suppose there is a topological obstruction in deforming between band insulators $a$ and $b$ without either closing the band gap or allowing symmetry breaking in the process. In equation, we may write $a \not \sim b$. However, it could be the case that the obstruction is resolved once we stack some additional band insulator $c$ with both $a$ and $b$, i.e., $a\oplus c \sim b \oplus c$ (Fig.\ 2a in the main text). From a physical point of view, $c$ may be an atomic insulator corresponding to some closed-shell electrons tightly localized to the underlying atoms. Since the deformation obstruction could be resolved by mixing with degrees of freedom that are buried deep below the Fermi energy, it is natural on physical grounds to disregard such apparent distinctions. Mathematically, one  say $a$ and $b$ are ``stably equivalent,'' and write $a \sim_{\rm s} b$.

The discussion above closely mirrors the key physical aspects of the notion of ``fragile'' vs. ``stable'' topology. Suppose a set of bands ${\rm FT}$ is topological in the sense that, for any trivial (i.e., atomic) band insulator ${\rm t}$, ${\rm FT} \not \sim {\rm t}$. Now we say ${\rm FT}$ has fragile topology if one can find some trivial ${\rm t'}$, ${\rm t''}$ such that ${\rm FT} \oplus {\rm t'} \sim {\rm t''}$ (Fig.\ 2b in the main text), and we say it is stably topological otherwise (Fig.\ 2c in the main text).

In closing, we note that the K-theory-based classifications of band insulators \cite{Kitaev, MooreFreed, Ktheory, Combinatorics,Gomi, Ken2017} are designed to (only) capture stable topological distinctions between band insulators, and therefore it does not automatically incorporate our present notion of triviality, which is defined with respect to Wannier representability. As an example, the number of occupied bands is a topological invariant in the K-theory sense, since there is no way to deform a set of $N$ bands into $N'$ bands when $N\neq N'
$. For physics applications, however, the electron filling in a band insulator is usually irrelevant to discussions for topological band insulators, as one can typically find a full set of atomic insulators which realizes all the possible band-insulator fillings
(We note in passing that, assuming either spinful or spinless fermions, the only exceptions to this rule (in the stable sense) among the 1,651 magnetic space groups were identified in Refs.\ \onlinecite{NC,MSG}.
These are instances where the stable band topology is manifested already in the electron filling.)
Our present notion of triviality of a band insulator, which is based on the existence of an atomic description, can be introduced into the K-theory framework by identifying the atomic-insulator subgroup in the full classification \cite{NC, MSG}.

\section{A more physical view on fragile topology
\label{app:PhysView}
}
While we have already provided a precise definition of fragile topology, it is helpful to shift the perspective from a more mathematical point of view to a physical one, concerned not with the topology of an isolated set of bands, but that of a band insulator:
 Consider placing the set of bands whose topology is to be determined at the bottom of the spectrum, and place the chemical potential above them so that they are the only filled bands. We will allow for the addition of {\it any} extra degrees of freedom {\it above} the Fermi level. This is rather physical---a bounded tight-binding model is only an approximation to any physical problem, and so it is unreasonable to forbid the addition of high-energy orbitals in the discussion. 
Now, we ask if we can tune some parameters and deform the system into an explicit atomic insulator, while preserving the band gap and symmetries throughout.
If yes, then we conclude the valence bands, our target set, are trivial; if no, then they are topological, and we have to further discern if the topology is stable or fragile. These two cases can be differentiated by further stacking with atomic insulators, corresponding to the addition of trivial bands below the Fermi level, and then ask if the new set of valence bands is adiabatically and symmetrically deformable into an atomic limit. 
We conclude our target set possesses fragile topology if and only if 
such a deformation is possible for some choice of additional trivial bands.

\section{Band representations
\label{app:BR}}
To be self-contained, we provide here a very brief introduction to the notion of ``band representations'' \cite{Zak1982, Bacry1988, Zak2000, Zak2001}. For details, please refer to Refs.\ \onlinecite{TopoChem,Graph, DBilbao}, and the references therein.

Loosely, a band representation is specified by two pieces of data: (i) the positions of the sites at which the electronic degrees of freedom reside, and (ii) how the local energy levels transform under the subgroup of the SG which leaves a site invariant.
In addition, band representations can be ``added,'' which corresponds physically to the stacking of the  local energy levels. Given a band representation, one can ask if it can be regarded as a stack of smaller ones, each involving fewer energy levels than the original. Whenever such an interpretation is possible, we say the band representation is ``composite.'' Any band representation which is not composite is called ``elementary.'' 

As defined, EBRs are the building blocks which generate all possible tightly localized atomic insulators under the stacking operation \cite{Bacry1988, Zak2000, Zak2001}.
Following the terminology in Refs.\ \onlinecite{TopoChem,Graph, DBilbao}, we refer to a time-reversal symmetric EBR as a ``physical'' EBR (PEBR). 
Note that all the PEBRs we discuss in this work are also EBRs \cite{TopoChem,Graph}, and hence we refrain from complicating the discussion by considering PEBRs in details.
Generally, it is possible that the energy bands corresponding to a (P)EBR can be split into valence and conduction bands separated by a band gap. When this is possible, we say the (P)EBR is decomposable \cite{TopoChem,Graph, DBilbao}. 

\end{document}